\documentclass[doublecol,linenumbers]{epl2}

\usepackage{graphicx}
\usepackage{subfigure}
\usepackage{amsthm}
\usepackage{amsmath}
\usepackage{amssymb}
\usepackage{verbatim}
\usepackage{dcolumn}
\usepackage{bm}
\usepackage{epsf}
\usepackage{color}
\usepackage[colorlinks=true,citecolor=blue,linkcolor=blue,urlcolor=blue]{hyperref}%
\usepackage{xcolor}
\usepackage{dsfont}

\newcommand{\la}{\left\langle}
\newcommand{\ra}{\right\rangle}

\newcommand{\e}[1]{\exp{\left(#1\right)}}

\newcommand{\bla}{bla\\bla\\bla\\bla\\bla}

\newcommand{\mc}[1]{\mathcal{#1}}

\newcommand{\mrm}[1]{\mathrm{#1}}

\title{Information scrambling -- a quantum thermodynamic perspective}

\author{Akram Touil\inst{1}  \and Sebastian Deffner\inst{2,3} \thanks{E-mail: \email{atouil@lanl.gov} and  \email{deffner@umbc.edu}}}
\shortauthor{Akram Touil and Sebastian Deffner}

\institute{\inst{1} Theoretical Division, Los Alamos National Laboratory, Los Alamos, New Mexico 87545\\
\inst{2} Department of Physics, University of Maryland, Baltimore County, Baltimore, MD 21250, USA\\
\inst{3} National Quantum Laboratory, College Park, MD 20740, USA
}

\abstract{Recent advances in quantum information science have shed light on the intricate dynamics of quantum many-body systems, for which quantum information scrambling is a perfect example. Motivated by considerations of the thermodynamics of quantum information, this perspective aims at synthesizing key findings from several pivotal studies and exploring various aspects of quantum scrambling. We consider quantifiers such as the Out-of-Time-Ordered Correlator (OTOC), the quantum Mutual Information, and the Tripartite Mutual Information (TMI), their connections to thermodynamics, and their role in understanding chaotic versus integrable quantum systems. With a focus on representative examples, we cover a range of topics, including the thermodynamics of quantum information scrambling, and the scrambling dynamics in quantum gravity models such as the Sachdev-Ye-Kitaev (SYK) model. Examining these diverse approaches enables us to highlight the multifaceted nature of quantum information scrambling and its significance in understanding the fundamental aspects of quantum many-body dynamics at the intersection of quantum mechanics and thermodynamics.}

\begin{document}

\maketitle

Quantum information scrambling, a dynamical phenomenon observed in complex quantum many-body systems, has garnered significant interest for its profound implications in various fields of modern research~\cite{Berry1989PS,Goldfriend2020PRE,Xu2020PRL,Xu2023arXiv}. At its core, scrambling is driven by the dispersion of initially localized information throughout a quantum system, leading to intricate patterns of entanglement and correlations. This concept not only challenges our understanding of quantum dynamics but also provides an informative window into the elusive nature of quantum chaos \cite{Xu2020PRL} and thermalization \cite{Swingle2018NP}.

The study of scrambling intersects with several critical areas of theoretical and experimental physics. For instance, in the realm of black hole physics, it offers insights into the information paradox and the nature of Hawking radiation \cite{calmet2022brief,antonini2018black}. In contrast, understanding scrambling in the context of quantum computing is essential for developing robust systems resistant to decoherence.

Motivated by thermodynamic considerations about processing information, this perspective aims at delving into the intricate details of quantum information scrambling, exploring its theoretical foundations, experimental verifications, and the multitude of its implications in modern physics. However, similar to all perspectives, the following exposition cannot serve as a comprehensive review, but only provides an introduction into the vast literature on the topic.

\section{From Black Holes to Quantum Information}

The exploration of quantum information scrambling has unveiled profound connections between the enigmatic nature of black holes and the fundamental principles of quantum physics. Black holes, traditionally studied within the realm of general relativity, have emerged as pivotal objects in understanding the quantum mechanical behavior of information in extreme gravitational fields. The seminal work by Hawking and others has led to the \emph{information paradox} in black holes, propelling the study of scrambling as a possible resolution \cite{Calmet2022EPL}.

Typically, the paradox is phrased as a question of whether or how information, that crosses the event horizon, is ``destroyed''. From the point of view of thermodynamics this question appears somewhat awkward, as it has been recognized that information is a thermodynamic resource that needs to be treated in complete analogy to heat and work \cite{Deffner2013PRX}. In other words, ``destroying information'' is thermodynamically equivalent to ``destroying energy''. Yet, it is a core principle of physics that energy cannot be created or destroyed, but only transformed into different forms.

Motivated by this insight, Hayden and Preskill \cite{Hayden2007JHEP} proposed a \emph{thermodynamically consistent approach} to resolving the black hole paradox. Their idea is that any information that crosses the event horizon of a singularity in space time, is rapidly and chaotically ``scrambled'' across this horizon. Eventually, the infalling information will be transformed into heat and re-emitted as Hawking radiation \cite{Almheiri2021RMP}. They further argued that if the black hole was a purely classical system, the event horizon would effectively act as an information  mirror. This means that any infalling information would rapidly be recoverable from the outgoing radiation. 

A more reasonable treatment requires considering the quantum entanglement between Hawking radiation and external observers. In other words, Hayden and Preskill \cite{Hayden2007JHEP} assert that black holes should be described as quantum communication channels. Consequently, \emph{quantum information scrambling} appears as \emph{the} thermodynamically motivated and consistent resolution of the black hole information paradox. Thus, one might expect that information scrambling was quickly picked up by the quantum thermodynamics community. Yet, Ref.~\cite{Hayden2007JHEP} was met with more immediate impact in cosmology \cite{Almheiri2021RMP} and quantum information theory \cite{Bao2021JHEP,Chen2022RP,Fisher2023}.

Additionally, theoretical advancements in string theory and the holographic principle suggest that the dynamics of black holes can be modeled by lower-dimensional quantum systems with high degrees of scrambling. This is exemplified by the Sachdev-Ye-Kitaev (SYK) model, a solvable model that exhibits maximal scrambling in the large $N$ limit, where $N$ is the total number of Majorana fermions in the system. More specifically, the SYK model is described by the Hamiltonian,
\begin{equation}
\label{eq:syk}
H_{\text{SYK}}=(i)^{\frac{q}{2}} \sum_{1 \leq i_{1}<i_{2}<\cdots<i_{q} \leq N} J_{i_{1} i_{2} \cdots i_{q}} \psi_{i_{1}} \psi_{i_{2}} \cdots \psi_{i_{q}},
\end{equation}
where $J_{i_{1} i_{2} \cdots i_{q}}$ are real independent random variables with values drawn from a Gaussian distribution with mean $\left\langle J_{i_{1} \cdots i_{q}}\right\rangle=0$ and variance $\left\langle J_{i_{1} \cdots i_{q}}^{2}\right\rangle=J^{2}(q-1) !/N^{q-1}$. The parameter $J$ (in the variance) sets the scale of the Hamiltonian. Further,  $\psi_{i}$ are Majorana field operators for $i\in \{ 1,\dots, N\}$. The chaotic dynamics of this model mimic the hypothesized behavior of quantum information in the vicinity of a black hole's event horizon.

Furthermore, the Anti-de Sitter/Conformal Field Theory (AdS/CFT) correspondence provides a framework for relating gravitational phenomena in black holes to quantum field theories invariant under conformal transformations. This duality offers a unique perspective on how quantum information might be scrambled in a black hole, encoded in the boundary CFT, and suggests that black holes could be the fastest scramblers in nature~\cite{Yasuhiro}.

\section{Quantifying Scrambling: The OTOC}

The natural question arises how to best quantify and characterize information scrambling and quantum chaos. In classical Hamiltonian dynamics, chaos can be identified from the exponential growth of the Poisson bracket \cite{Goldstein2002}. Hence, it may not be a surprise that the most prominent tool to diagnose scrambling of quantum information is a closely related quantity -- the out-of-time-ordered correlation function (OTOC) \cite{Swingle2018NP}.

The OTOC is a four-point correlation function that measures the growth of operators in the Heisenberg picture, defined as,
\begin{equation}
\label{eq:OTOC}
F(t) = \left\langle W^\dagger(t) V^\dagger W(t) V \right\rangle,
\end{equation}
where $ W(t)\equiv \e{i/\hbar\,Ht}\, W \,\e{-i/\hbar\,Ht}$, and $H$ is the Hamiltonian of the system of interest. The average, i.e. expectation value, is taken over the initial state of the system. It is often convenient to also analyze the expectation value of the squared commutator \cite{Roberts2016PRL,Swingle2018NP},
\begin{equation}
\label{eq:C}
 C(t)\equiv\la \left[W^\dagger(t), V\right]^2\ra=2 \left(1-\mathfrak{R}(F(t)\right)\,,
\end{equation}
where $\mathfrak{R}(F(t)$ denotes the real part of the OTOC \eqref{eq:OTOC}.

For many-body quantum systems, $W$ and $V$ have to be chosen as meaningful and accessible observables reflecting the complexity of the system. However, the quantum-classical analogy becomes much more stringent if we set $W=x$ as position and $V=p$ as momentum. In this case, $C(t)$~\eqref{eq:C} is nothing but the quantum version of the average Poisson bracket. Hence, quantum information scrambling is evidenced by any growth of $C(t)$, whereas an exponential scaling of $C(t)$ as a function of time signifies quantum chaotic dynamics.

The OTOC was not actually defined for information scrambling, but rather first appeared in the context of spin echos \cite{Hahn1950PR}. However, the OTOC has proven itself to be uniquely suited to analyze the growth of correlations in many-body systems~\cite{Swingle2016PRA,Hashimoto2017JHEP,Swingle2017PRB,Swingle2018PRA,Khemani2018PRX,Xu2020NP,Akutagawa2020JHEP,Belyansky2020PRL,Garcia2021PRR,Braumuller2022NP}. Moreover, it is an experimentally accessible quantity, which has been exploited to demonstrate information scrambling in, e.g., ion traps \cite{Landsmann2019Nature}.

\section{Experimental Verification of Scrambling}

Recent years have seen a surge in experimental efforts exploring fundamental concepts of quantum thermodynamics. To date, experiments have been reported employing, e.g., optomechanical systems, nuclear magnetic resonance, nitrogen vacancy centers, superconducting systems, and many more. For a comprehensive exposition we refer to a recent review article \cite{Myers2022AVSQS}. However, owning to their exquisite controllability, ion traps are arguably the most important platform for quantum thermodynamics, with which many of the first experiments have been realized~\cite{Huber2008PRL,Abah2012PRL,An2015NP,Rossnagel2016Science,Smith2018NJP}.

Therefore, for the present perspective the experiment by Landsman \etal \cite{Landsmann2019Nature} is of particular importance. Their experiment was motivated by the fact that the decay of the OTOC \eqref{eq:OTOC} does not uniquely signify scrambling, but could also be due to decoherence. Following this motivation, in the next sections, we will summarize the main results of our own comprehensive analysis of scrambling in decohering environments \cite{Touil2021PRXQ} from a thermodynamic perspective. Landsman \etal \cite{Landsmann2019Nature} based their work entirely on the experimentally accessible OTOC.

It had been shown by Yoshida and Yao \cite{Yoshia2019PRX} that the Hayden-Preskill protocol \cite{Hayden2007JHEP} for scrambling in black holes can be mapped onto a problem of quantum teleportation. In fact, maximally scrambling dynamics is necessary for successful teleportation. Therefore, the teleportation fidelity provides an independent diagnostics for scrambling. Landsman \etal \cite{Landsmann2019Nature} implemented the corresponding quantum circuit implementable in their seven-qubit quantum computer, which is realized in a crystal of trapped $^{171}$Yb$^+$. This experiment provided convincing evidence of information scrambling throughout the quantum computer, even in the presence of decoherence. However, it also highlighted the shortcomings of the OTOC as a sole identifier of scrambling, which made the quest for other quantifiers instrumental.

\section{Other Quantifiers of Scrambling}

To date, the OTOC has remained the most important quantifier of scrambling in experiments. Further aspects of scrambling dynamics were elucidated in ion traps \cite{Garttner2017NP,Joshi2020PRL,Green2022PRL}. In addition,  information scrambling has also been demonstrated on Google's Sycamore chip \cite{Mi2021Science,Jafferis2022Nature}, IBM's Q Experience \cite{Harris2022PRL}, a ladder-type superconducting \cite{Zhu2022PRL} and superconducting qutrits \cite{Blok2021PRX} quantum processors, and in NMR \cite{Li2017PRX}. Moreover, experiments have been proposed for macroscopic spins \cite{Blocher2022PRA}, quantum many-body spin models \cite{Joshi2022PRX,Sundar2022NJP}, and neutral atom arrays \cite{Hashizume2021PRL}. Yet, also several other metrics have been developed to capture different aspects of this complex phenomenon. These quantifiers provide additional insights into the dynamics of entanglement, chaos, and information dispersal~\cite{Yunger2018PRA,Iyoda2018PRA,Seshadri2018PRE,Vermersch2019PRX,Alonso2019PRL,Yunger2019CP,Yan2020PRL,Touil2020QST,Cao2022PRR,Zonnios2022PRL,Bhattacharyya2022,Omanakuttan2023PRA}.

\begin{figure}
	\centering 
	\includegraphics[width=.48\textwidth]{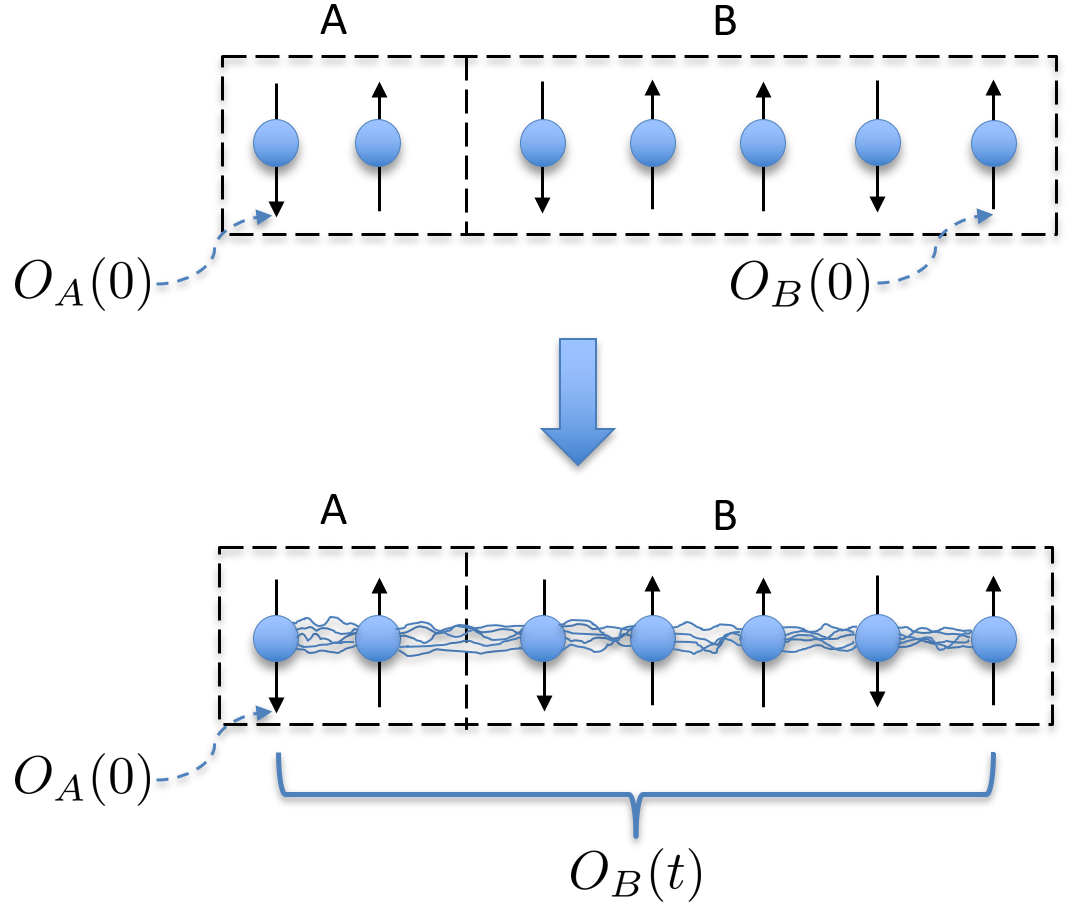}
	\caption{Depiction of the underlying substructure illustrating the support for the local operators $O_A$ and $O_B$, within a specific spin chain. Following the application of a scrambling unitary, $O_B$ evolves into a non-local operator, expanding into the support of $A$, thereby initiating the decay of the OTOC~\eqref{eq:OTOC}.}
	\label{chain}
\end{figure}

The main conceptual problem of using the OTOC as a scrambling quantifier stems from the seemingly arbitrary choice of operators $V$ and $W$. For bipartite quantum systems, $A\otimes B$, it is most convenient to consider observables that intially only have support on either $A$ or $B$, and hence $V=O_{A}$ and $W=O_{B}$. In Fig.~\ref{chain} we illustrate the ensuing scrambling of information.

To further alleviate the arbitrariness, one typically considers a Haar average over all possible observables with such support. Interestingly, for quantum systems comprised of qubits, such as the experimental system analyzed in Ref.~\cite{Landsmann2019Nature}, the Haar average is equivalent to an average over the Pauli group for each operator. Note, however, that in practice, an experimentalist has only access to a set of measurements they can implement depending on the physical setup. Hence, average OTOC are mostly of theoretical interest, as they  do not describe what can be implemented in an experiment. Therefore, constructing alternative measures of scrambling appears highly desirable.

\subsection{Mutual information}
As a staple of information theory, the mutual information seems to be the natural quantifier of scrambling in many-body systems. It is given by,
\begin{equation}
\label{eq:mut_info}
\mathcal{I}(t) = \mathcal{S}_{A}(t) + \mathcal{S}_{B}(t) - \mathcal{S}_{S}(t),
\end{equation}
where $S_{i} = -\mathrm{tr}(\rho_{i}\ln(\rho_{i}))$ represents the von Neumann entropy of system $i$ with the density matrix $\rho_{i}$. 

In Ref.~\cite{Touil2020QST}, we showed that
\begin{equation}
\label{eq:main1}
\mathcal{I}(t) \geq \bar{\mathcal{O}}(0) - \bar{\mathcal{O}}(t)\,,
\end{equation}
where $\bar{\mathcal{O}}(t) \equiv \langle O_{A}O_{B}(t)\,O_{A}O_{B}(t)\rangle_{\mathrm{avg}}$ is the Haar average of the OTOC. Equation~\eqref{eq:main1} holds true for all times $t > 0$, and shows that the decay of the OTOC $\bar{\mathcal{O}}(t)$ is upper bounded by the growth of mutual information.

Interestingly, Eq.~\eqref{eq:main1} provides a mathematically rigorous statement of the intuitive understanding of scrambling, which is often equated with the augmentation of bipartite correlations between subsystems $A$ and $B$. Namely, the deacy of the OTOC is governed by the growth of correlations in a complex quantum many body system.

Even more importantly, the quantum mutual information is directly related to stochastic thermodynamic quantities, which offers a lucid thermodynamic description of scrambling. More specifically, we showed~\cite{Touil2020QST} that
\begin{equation}
\mathcal{\dot{I}} \leq \mathcal{A}\, \dot{\mathcal{S}}_A + \mathcal{B}\, \dot{\mathcal{S}}_B + \mathcal{C}\,|\dot{\mathcal{S}}_E|\,,
\label{8}
\end{equation}
where $\mathcal{A}$, $\mathcal{B}$, and $\mathcal{C}$ represent discrete analogs of the Frieden integral, as referenced in~\cite{nikolov1994limitation,brody1995upper}, and are dependent solely on the spatial geometry of the problem. Additionally, $\dot{\mathcal{S}}_A$ and $\dot{\mathcal{S}}_B$ denote the stochastic irreversible entropy production within subsystems $A$ and $B$, as discussed in \cite{esposito2010three}. Lastly, $\dot{\mathcal{S}}_E$ refers to the flow of information between $A$ and $B$.

\begin{figure*}
\begin{minipage}[l]{.48\textwidth}
Decoherence in computational basis\\
\vspace{1.2em}
\includegraphics[width=8.5cm]{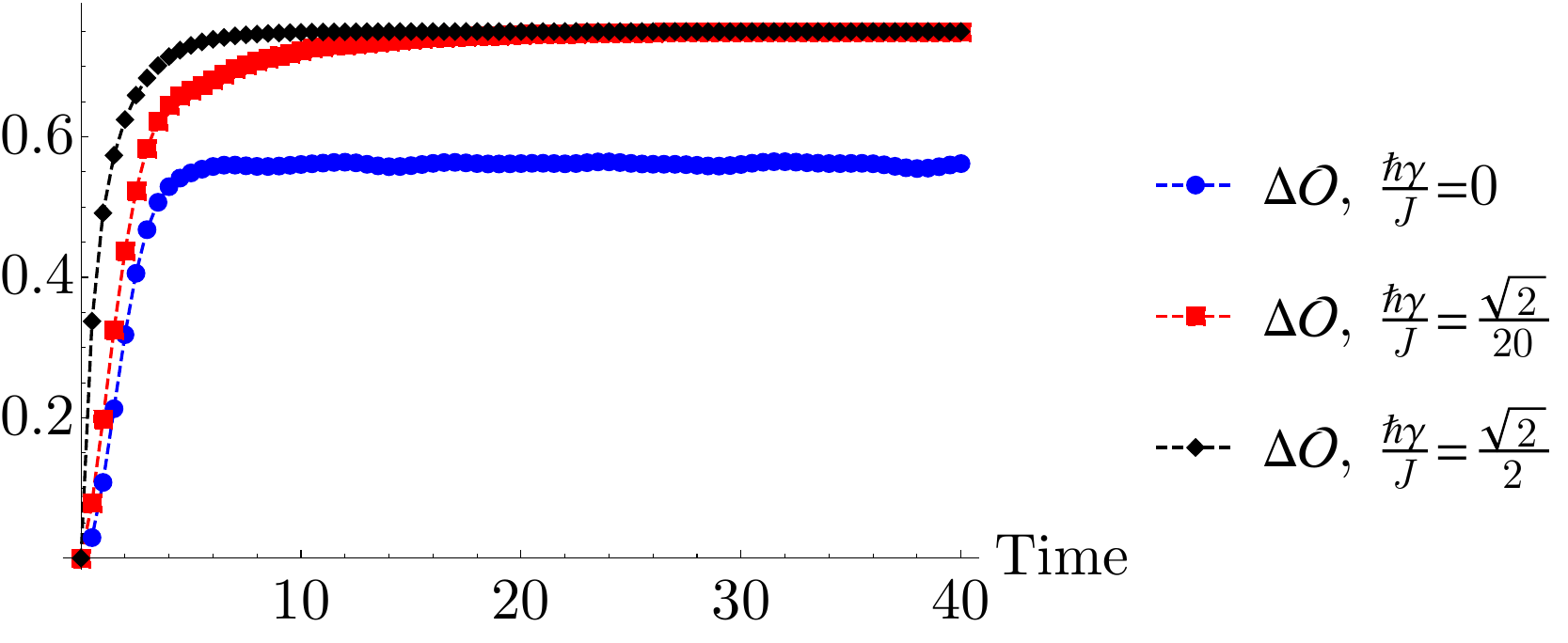}
\end{minipage}
\hfill
\begin{minipage}[l]{.48\textwidth}
Decoherence in energy basis\\
\vspace{1.2em}
\includegraphics[width=8.5cm]{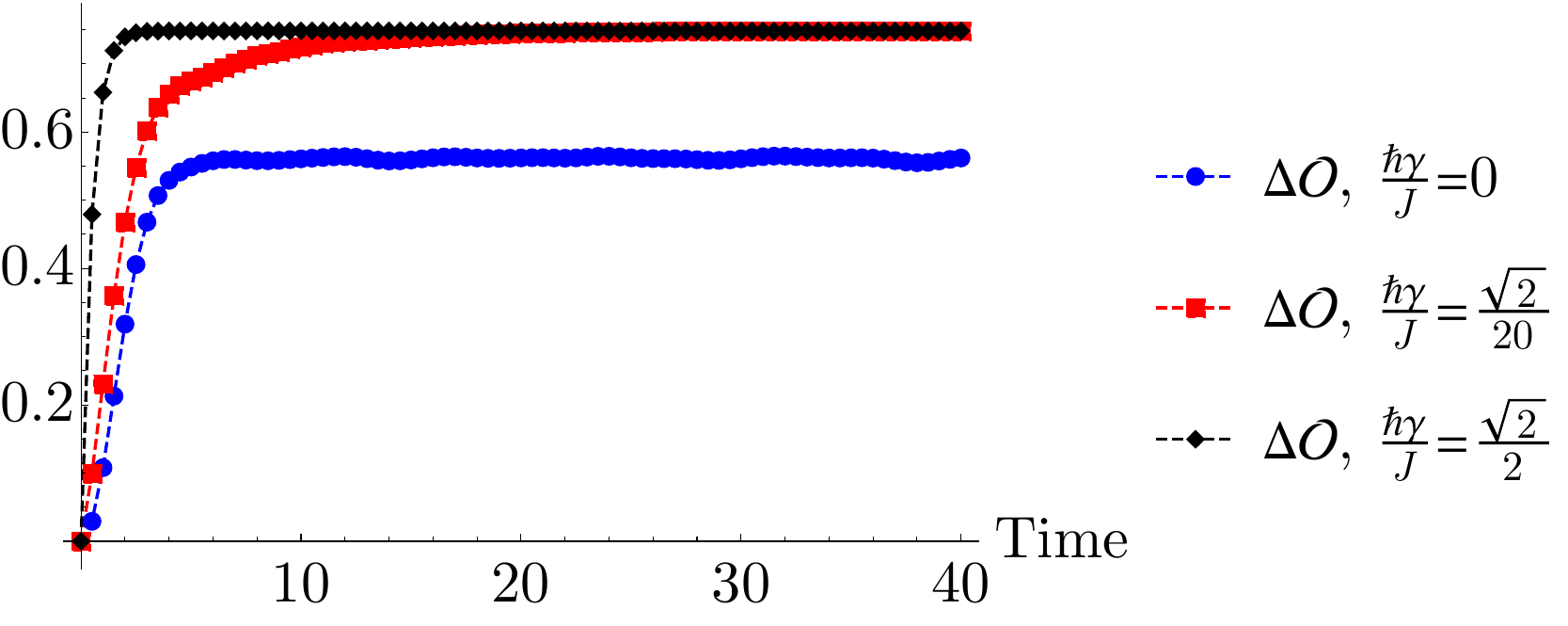}
\end{minipage}
\caption{Plots of the growth of the average OTOC as a function of time for the SYK-model with $N=12$ and an initial ``all-up'' state: $|0...0\rangle$. Plots were generated after averaging over $10^2$ realizations.}
	\label{otoc}
\end{figure*}

\subsection{Quantum work statistics}

In fact, understanding the thermodynamics of scrambling is under active investiagtion in the literature \cite{Yunger2017PRA,Campisi2017PRE,Chenu2018SR,Knap2018PRB,Chenu2019Quantum,Styliaris2021PRL,Mikkelsen2022PRL,Ohanesjan2023JHEP}. 

In particular, the two-point measurement scheme \cite{Talkner2007PRE}  of quantum work is directly related to the OTOC~\cite{Campisi2017PRE}. By expressing one of the OTOC operators $V$ in exponential form, $V=e^{i u O}$ with an appropriate hermitian operator $O$ and real number $u$. Reference~\cite{Campisi2017PRE} proposed the wing-flap protocol: (i) prepare the system in some state $\rho$; (ii) measure $O$; (iii) evolve the system with $H$ for a time $t=\tau$; (iv) apply the wing-flap perturbation $W$; (v) evolve the system with $-H$ for a time $t=\tau$; (vi) measure $O$.

The projective measurements of $O$ yield eigenvalues $O_n$ and $O_m$, correspondingly collapsing the system into eigenstates $|n\rangle$ and $|m\rangle$. It is the easy to see \cite{Campisi2017PRE} that the characteristic function, that is the Fourier transform of the probability density function of $O_m-O_n$ is identical to the OTOC.

This close relationship between the quantum work statistics and the OTOC becomes even more stringent, once one recognizes that for sudden changes the characteristic function becomes a Loschmidt echo~\cite{Silva2008PRL}. More generally, in Ref.~\cite{Yan2020PRL}, it was proven that the \emph{thermal Haar average} of the OTOC (for infinite temperature) is equal to the thermal average of the Loschmidt echo. Namely, we have
\begin{equation}
\overline{F_{\beta=0}(t)} \approx \overline{\left|\left\langle e^{i\left(H_B+V_\alpha\right) t} e^{-i\left(H_B+V_{\alpha^{\prime}}\right) t}\right\rangle_{\beta=0}\right|^2},
\end{equation}
where $V_{\alpha}$ are noise operators, and the average is over all noise realizations, and $H_B$ is the self-Hamiltonian on the support $B$.

\subsection{Tripartite Mutual Information (TMI)}

We conclude this section with another variation of the mutual information that involves three supports, namely the Tripartite Mutual Information (TMI), which offers a way to measure the distribution of information among multiple subsystems. The TMI for subsystems $A$, $B$, and $C$ in a quantum state $\rho$ is defined as,
\begin{equation}
    \mc{I}_3(A : B : C) = \mc{I}(A : B) + \mc{I}(A : C) - \mc{I}(A : B C),
\end{equation}
where $\mc{I}(X : Y)$ denotes the mutual information between subsystems $X$ and $Y$ \eqref{eq:mut_info}. A negative value of TMI indicates that quantum information initially localized in one subsystem has been effectively scrambled across the entire system. This quantity was studied extensively in Ref.~\cite{Iyoda2018PRA}.

\section{Scrambling in Open Systems}

Analyzing quantum information scrambling in open systems presents formidable challenges compared to closed systems. In open quantum systems, interactions with the environment play a crucial role in the dynamics. This interaction often leads to decoherence, which can either facilitate or hinder the scrambling process. A number of studies tackled this question~\cite{Yoshia2019PRX,Xu2019PRL,Touil2021PRXQ,Zanardi2021PRA,Xu2021PRB,Dominguez2021PRA,Cornelius2022PRL,Han2022Entropy,Andreadakis2023PRA} from various perspectives.

One of the key models used to study scrambling in open systems is the Lindblad master equation, which describes the time evolution of the density matrix $\rho$ of an open quantum system,
\begin{equation}
    \frac{d\rho}{dt} = -i[H, \rho] + \sum_{k} \left( L_k \rho L_k^\dagger - \frac{1}{2} \{L_k^\dagger L_k, \rho\} \right),
\end{equation}
where $H$ is the Hamiltonian of the system, and $L_k$ are the Lindblad operators that model the system's interaction with its environment. The first term on the right-hand side represents the unitary evolution, while the sum accounts for dissipation into the environment.

Scrambling in open systems can be characterized by the decay of quantum correlations and coherences. For example, the behavior of the OTOC in an open system can reveal how environmental interactions affect the rate and nature of scrambling. Additionally, the entanglement dynamics in open systems, as quantified by entanglement measures like concurrence or entanglement entropy, provide valuable insights into the scrambling process under the influence of external noise and decoherence.

In Ref.~\cite{Touil2021PRXQ} we analyzed the effects of decoherence in both energy and computational bases. In Fig.~\ref{otoc} we present the Haar averaged OTOC \eqref{eq:OTOC} for the SYK model \eqref{eq:syk} for different coupling intensities between $\mathcal{S}$ and $\mathcal{E}$, specifically at zero ($\hbar\gamma/J = 0$), weak ($\hbar\gamma/J \ll 1$), and moderate ($\hbar\gamma/J \simeq 1$) levels. Here, the parameter $\gamma$ is the strength of the interaction with the environment.

Note that maximal scrambling is indicated by a maximum of the mutual information, which for qubit models simply becomes $\mathcal{I} = 2 \ln(2)$; therefore, any environmental impact causes $\mathcal{I}$ to decrease below $2 \ln(2)$ over extended periods. Hence, it appears natural to consider the dynamics of  $\mathcal{I}$. To this end, we showed in Ref.~\cite{Touil2021PRXQ} that in open systems the change of the von Neumann entropy can be separated into three terms,
\begin{equation}
\mathcal{I(\mathcal{S}:\mathcal{E})}+\Delta S_\mrm{ex}+D(\rho_{\mathcal{E}} || \rho^\mrm{eq}_{\mathcal{E}} )=\Delta S_\mc{S}\,.
\end{equation}
This equation expresses that any departure from ideal unitary scrambling can be attributed to three factors: (i) the development of correlations between $\mc{S}$ and $\mc{E}$, (ii) the thermal interchange between $\mc{S}$ and $\mc{E}$, and (iii) the deviation of $\mc{E}$ from thermal equilibrium. 

\begin{figure*}
\begin{minipage}[l]{.48\textwidth}
Decoherence in computational basis\\
\vspace{.5em}
\includegraphics[width=8.5cm]{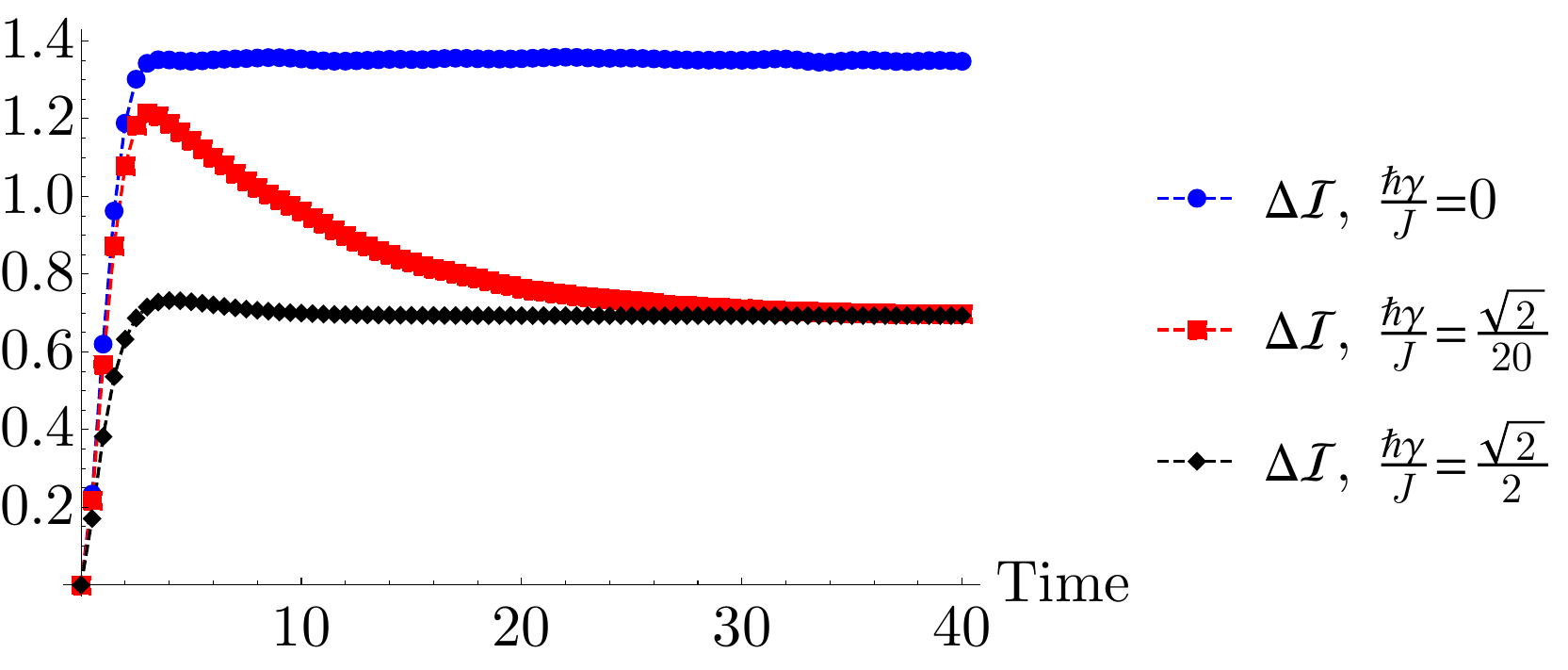}
\end{minipage}
\hfill
\begin{minipage}[l]{.48\textwidth}
Decoherence in energy basis\\
\vspace{.5em}
\includegraphics[width=8.5cm]{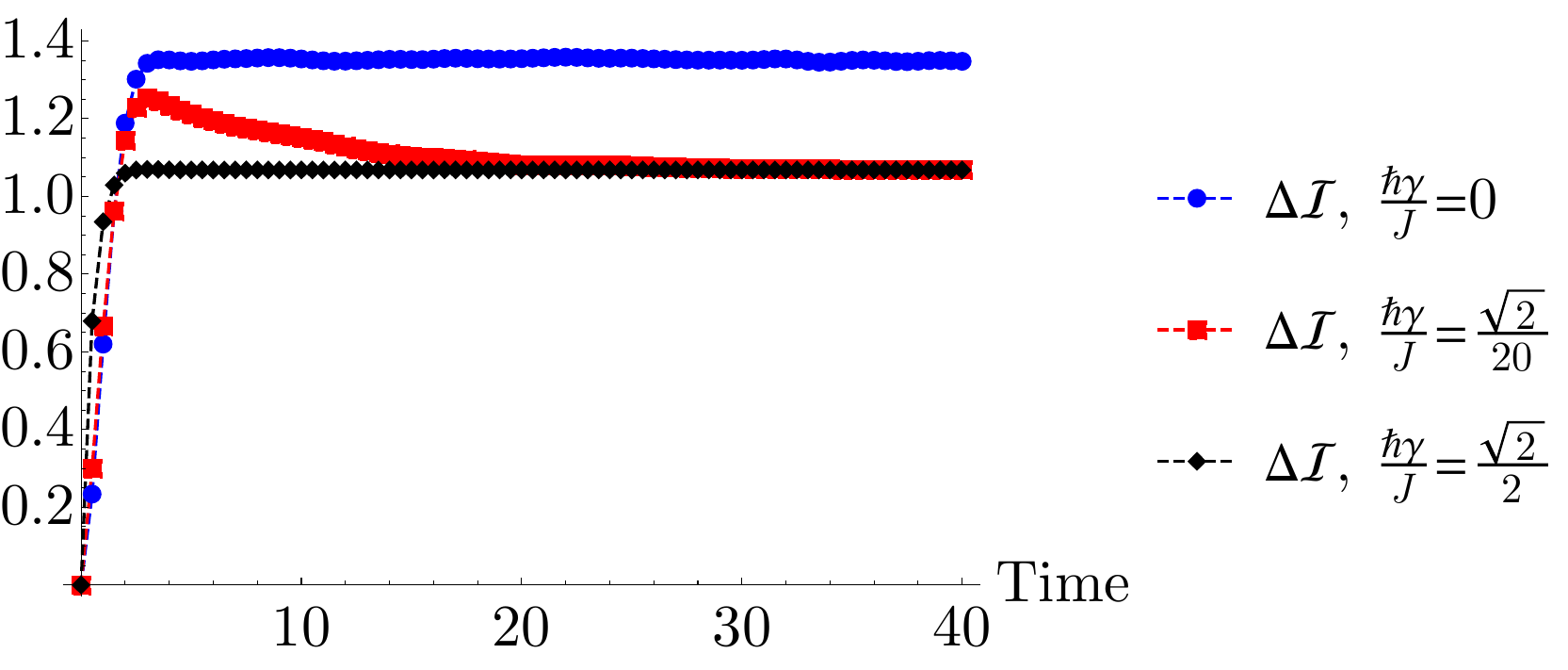}
\end{minipage}
\caption{Plots of the evolution of the change of mutual information \eqref{eq:mut_info} for the SYK-model \eqref{eq:syk} with $N=12$, for an initial ``all-up'' state.  Plots were generated after averaging over $10^2$ realizations.}
	\label{full_illust1}
\end{figure*}

In Fig.~\ref{full_illust1} this insight is further illustrated for the SYK model as a chaotic model and the LMG model as a fully integrable model, whose Hamiltonian reads
\begin{equation}
H_{\mathrm{LMG}}=-\frac{J}{N} \sum_{i<j}^{N}\left(\sigma_{x}^{i} \sigma_{x}^{j}+\sigma_{y}^{i} \sigma_{y}^{j}\right)-\sum_{i=1}^{N} \sigma_{z}^{i}\,.
\label{lmgh}
\end{equation}
In Ref.~\cite{Touil2021PRXQ}, we further found that the mutual information exhibits universal behavior. Any deviation from the monotonic growth of the mutual information points to some interaction with the environment, given that the unitary dynamics are indeed scrambling. This interaction might be in the form of pure decoherence (destruction of coherences + no dissipation), or destruction of coherences accompanied with dissipation.

Additionally, it is important to mention that a specific modification of the OTOC was defined in Ref.~\cite{Zanardi2021PRA}. This modification, dubbed ``open bipartite OTOC'', essentially captures the operator entanglement of two operators that start (at $t=0$) in different supports. More specifically, this paper adopts the Heisenberg picture, 
where the evolution of observables is modeled by a Completely Positive (CP) map, 
\(\mathcal{E}\). A quantum channel \(\mathcal{E}\) is considered unital if it leaves 
the maximally mixed state, represented by \(\frac{\mathbb{I}}{d}\), unchanged. Such 
channels include unitary evolution, projective measurements without postselection, 
and dephasing channels. A quantum channel is trace-preserving if its adjoint 
\(\mathcal{E}^{\dagger}\) is unital. For simplicity, the paper assumes 
\(\mathcal{E}^{\dagger}\) to be unital, implying \(\mathcal{E}\) is a quantum channel.

The open bipartite Out-of-Time-Order Correlator (OTOC) is defined as~\cite{Zanardi2021PRA}, 
\begin{equation}
F(t)=\frac{1}{2d} \mathbb{E}_{V_A, W_B}\left\|\left[\mathcal{E}\left(V_A\right), W_B\right]\right\|_2^2\,,
\end{equation}
where the operators are averaged using the Haar measure of unitaries discussed above. From this form, this modified OTOC captures bipartite correlations between the supports $A$ and $B$.

The numerical results were identical to what the quantum mutual information captures, which is not surprising given the parallel between the mutual information and the OTOC, as shown in the previous section of this perspective, and detailed in Refs.~\cite{Touil2020QST,Touil2021PRXQ,touil2023black}.

\begin{figure*}
\begin{minipage}[l]{.48\textwidth}
Decoherence in computational basis\\
\vspace{.5em}
\includegraphics[width=8.5cm]{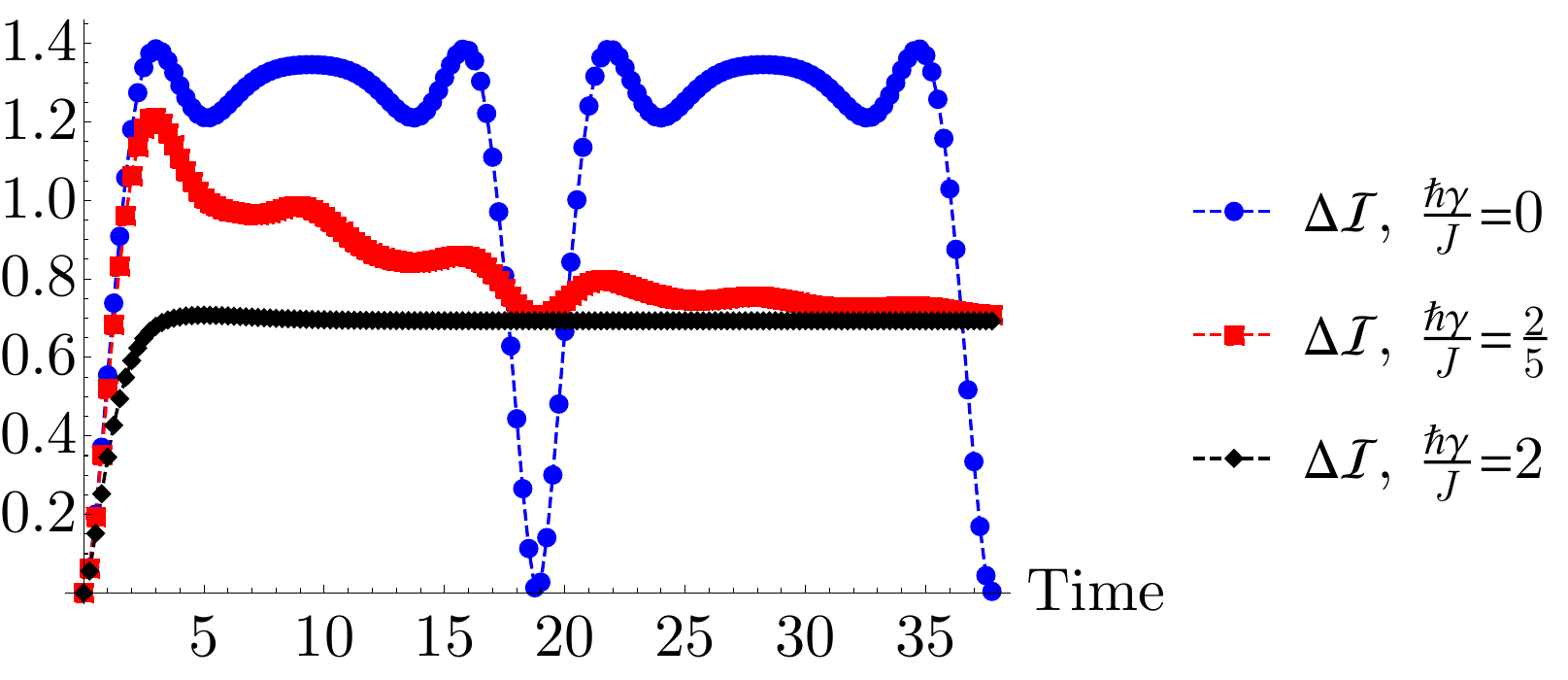}
\end{minipage}
\hfill
\begin{minipage}[l]{.48\textwidth}
Decoherence in energy basis\\
\vspace{.5em}
\includegraphics[width=8.5cm]{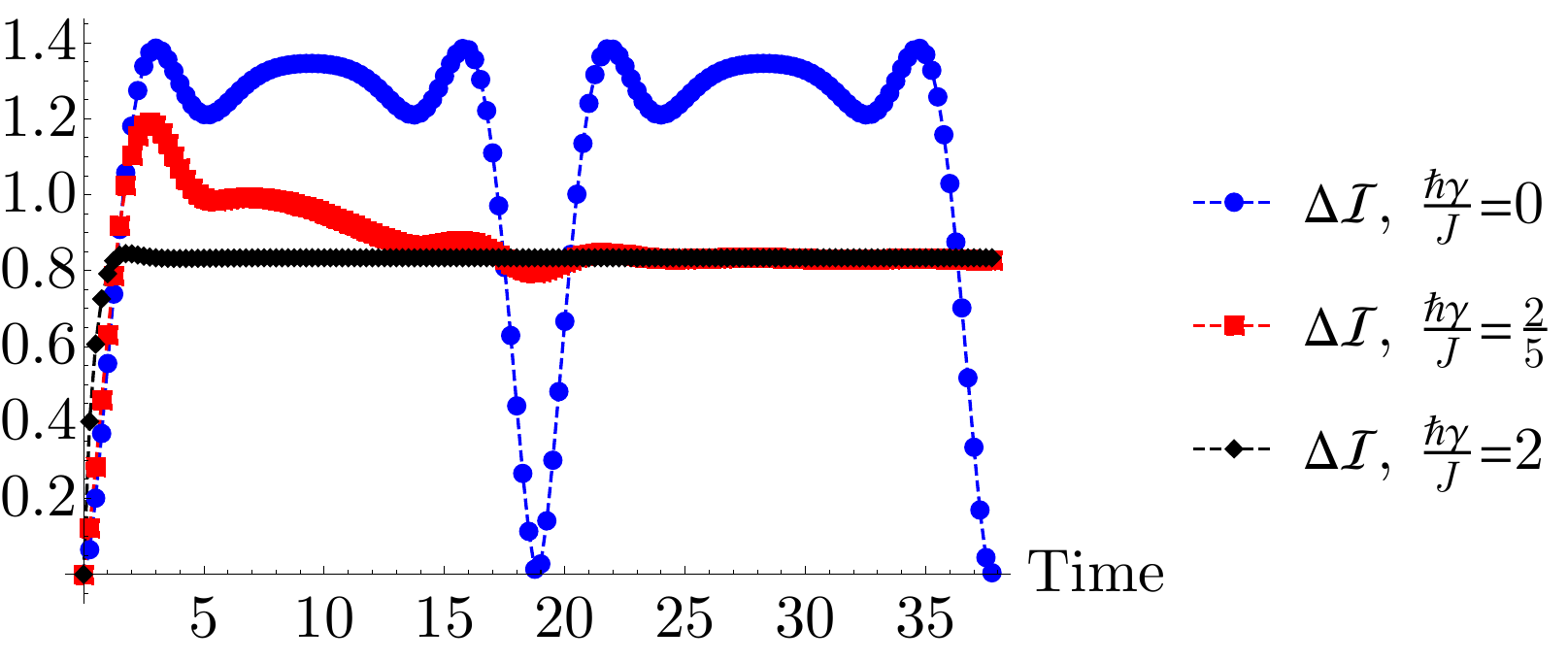}
\end{minipage}
	\caption{Plots of the evolution of the change in mutual information \eqref{eq:mut_info} for the LMG-model with $N=6$ and an initial N\'eel state.}
	\label{full_illust2}
\end{figure*}

\section{Applications and Future Directions}

We close this perspective with a few considerations on future directions. In fact, the study of quantum information scrambling has far-reaching implications and applications, extending across various domains of physics and opening new avenues for future research. As our understanding of scrambling, from a quantum thermodynamic perspective, deepens, its potential impact on several fields becomes increasingly evident~\cite{Carvelli2020PRR,Yuan2022PRR,Garcia2022JHEP,Bin2023PRB,Li2023Science,Garcia2023PNAS}.

\subsection{Quantum Computing and Error Correction}
In the realm of quantum computing, scrambling plays a crucial role in understanding and mitigating quantum errors. Quantum error correction codes can be analyzed through the lens of scrambling, where the ability of a system to disperse local errors throughout its many degrees of freedom can be harnessed for more robust error correction schemes. For example, the surface code, a widely studied error correction code, can be analyzed using scrambling metrics to optimize its fault-tolerance,
\begin{equation}
    \text{Surface Code Hamiltonian: } H_{\text{SC}} = -\sum_{s} A_s - \sum_{p} B_p,
\end{equation}
where $A_s$ and $B_p$ are stabilizer operators corresponding to the vertices (stars) and plaquettes of the lattice, respectively. This hints at direct applications of information scrambling in cybersecurity for quantum computers~\cite{raheman2022future}.

\subsection{Many-Body Localization}
Another exciting avenue is the exploration of many-body localization (MBL) in the context of scrambling. MBL systems, characterized by the absence of thermalization, present a unique platform to study how scrambling dynamics are altered in the presence of disorder and localization,
\begin{equation}
    H_{\text{MBL}} = \sum_{i} h_i n_i + \sum_{i < j} J_{ij} n_i n_j,
\end{equation}
where $n_i$ are the occupation numbers, $h_i$ represents on-site potentials, and $J_{ij}$ are interaction strengths~\cite{chen2016universal}.

\subsection{Quantum Metrology}
Quantum scrambling also finds applications in quantum metrology, enhancing the precision of quantum measurements. The sensitivity of quantum systems to initial conditions, as quantified by scrambling metrics, can be exploited to improve the accuracy of quantum sensors and clocks~\cite{BinNik}.

\subsection{Future Theoretical and Experimental Challenges}
Looking ahead, both theoretical and experimental challenges remain in fully understanding and harnessing quantum scrambling. Experimentally, realizing systems that can coherently demonstrate scrambling dynamics over extended periods is a key challenge. Theoretically, developing a unified framework that incorporates scrambling in diverse quantum systems remains an active area of research.

In conclusion, the applications and future directions of quantum information scrambling span a wide range of fields, from quantum computing and black hole physics to condensed matter and quantum metrology. As research in this area continues to evolve, it holds the promise of revealing deeper truths about the quantum world and unlocking new technological capabilities in quantum science.

\section{Concluding Remarks}

As we conclude this perspective on quantum information scrambling, it is evident that this phenomenon has become crucial in our understanding of quantum mechanics, with broad implications across various fields of physics. From providing insights into the fundamental nature of black holes to shaping the development of quantum computing, scrambling has emerged as a key concept in both theoretical and experimental physics.

In summary, this perspective highlights how quantum information scrambling is key in our effort to understand the quantum world. Its study tests our current knowledge and leads to new opportunities in technology and basic quantum science. As we continue to explore this area of research from a thermodynamic viewpoint, we expect to gain more understanding of our quantum universe and its different applications.

\acknowledgments

A.T. acknowledges support from the U.S DOE under the LDRD program at Los Alamos. S.D. acknowledges support from  the U.S. National Science Foundation under Grant No. DMR-2010127 and the John Templeton Foundation under Grant No. 62422.

\bibliographystyle{eplbib}
\bibliography{epl_scramb}
\end{document}